\documentstyle[multicol,floats,aps,prb]{revtex}
\begin{document}
\draft

\title{Radial marginal perturbation of two-dimensional systems and conformal
invariance}

\author{Lo\"{\i}c Turban}

\address{Laboratoire	de Physique du Solide, Universit\'e de Nancy I, 
Bo\^\i te Postale 239,\\ F54506 Vand\oe uvre l\`es Nancy CEDEX, France\\
{\rm (Received 31 May 1991)}}

\maketitle

\begin{abstract}
The conformal mapping $w=(L/2\pi)\ln z$ transforms the critical plane with a
radial perturbation $\alpha\rho^{-y}$ into a cylinder with width $L$ and a
constant deviation $\alpha(2\pi/L)^y$ from the bulk critical point when the
decay exponent $y$ is such that the perturbation is marginal. {}From the known
behavior of the homogeneous off--critical system on the cylinder, one may
deduce the correlation functions and defect exponents on the perturbed plane.
The results are supported by an exact solution for the Gaussian model.
\end{abstract} 

\vglue.5truecm

\begin{multicols}{2}
\narrowtext

Although conformal invariance requires, in principle, translation and rotation
invariance besides global scale invariance,\cite{belavin84} some progress has
been made in recent years concerning the behavior of critical two-dimensional
systems with defects. The case of line defects with commensurate
configurations and extended defects has been extensvely studied,$^{2-6}$ the
gap-exponent relations have been verified, and towerlike spectra has been
obtained for marginal perturbations decaying as a power of the distance from
either a free surface,\cite{igloi90a} the Hilhorst--van Leeuwen
model,\cite{hilhorst81} or a line.\cite{igloi90b} In this work we present
results on marginal radial perturbations of critical systems which are likely
to be applicable to any conformally invariant model in two dimensions.

Let us consider a perturbed critical system on the plane $(\rho,\theta)$ with
a free-energy density functional 
\begin{equation}
F[\varphi]=F_c[\varphi]+\int\!\!\rho\, d\rho\, d\theta\,
\alpha\rho^{-y}\,\chi(\rho,\theta)\; ,
\label{e1}
\end{equation}
where the deviation from the critical point in the last term decays as a power
of the distance from the origin with an amplitude
$\Delta_\chi(\rho)=\alpha\rho^{-y}$. Through the conformal transformation
$w=(L/2\pi)\ln z$, the plane is mapped onto a cylinder $(u,v)$ such that
\begin{mathletters}
\label{e2}
\begin{eqnarray}
u&=&(L/2\pi)\;\ln\rho\; ,\qquad-\infty<u<+\infty\; ,\\
v&=&L\theta/2\pi\; ,\qquad0\leq v<L\; ,
\end{eqnarray}
\end{mathletters}
with a dilatation factor:
\begin{equation}
b(z)=\vert w'(z)\vert^{-1}=2\pi\rho/L\; .
\label{e3}
\end{equation}
The radial inhomogeneity transforms according to 
\begin{equation}
\Delta_\chi(w)=b(z)^{y_\chi}\alpha\rho^{-y}=\alpha(2\pi/L)^{y_\chi}
\rho^{y_\chi-y}\; .
\label{e4}
\end{equation}
The perturbation is marginal when $y=y_\chi$ since in a global transformation 
$\alpha$ is changed into $b^{y_\chi-y}\alpha$.\cite{cordery82} In the
following we shall only consider the marginal case, assuming that conformal
invariance remains valid. The deviation from criticality on the cylinder is
then a constant
\begin{equation}
\Delta_\chi=\alpha(2\pi/L)^{y_\chi}\; ,
\label{e5}
\end{equation}
preserving translation invariance in both directions. The two-point correlation
function for an operator $\phi$ with scaling dimension $x_\phi$ transforms as
\cite{belavin84}
\begin{equation}
\langle\phi(w_1)\phi(w_2)\rangle=\vert w'(z_1)\vert^{-x_\phi}\vert
w'(z_2)\vert^{-x_\phi}\langle\phi(z_1)\phi(z_2)\rangle\; ,
\label{e6}
\end{equation}
so that the correlation function on the perturbed critical plane may be deduced
from the known behavior of the system on the off-critical cylinder,\cite{kogut79}
\begin{eqnarray}
&&\langle\phi(u_1,v_1)\phi(u_2,v_2)\rangle\nonumber\\
&&\qquad=\sum_{\alpha\neq0}\vert\langle0\vert\phi\vert\alpha\rangle\vert^2
\exp[-(E_\alpha-E_0)(u_2-u_1)\nonumber\\
&&\qquad\qquad\qquad\qquad\qquad\quad\!+iP_\alpha(v_2-v_1)]\; ,
\label{e7}
\end{eqnarray}
where $E_\alpha$ and $P_\alpha$ are the energy and momentum eigenvalues associated
with the eigenstate $\vert\alpha\rangle$ of the transfer operator
$\underline{T}=e^{-\tau\underline{H}}$ and $\vert0\rangle$ is the ground state of
the Hamiltonian $\underline{H}$. Collecting these results, one obtains
\begin{eqnarray}
&&\langle\phi(\rho_1,\theta_1)\phi(\rho_2,\theta_2)\rangle\nonumber\\
&&\qquad=(L/2\pi)^{2x_\phi}\sum_{\alpha\neq0}\vert\langle0\vert\phi
\vert\alpha\rangle\vert^2\rho_1^{-x_\phi+(L/2\pi)(E_\alpha-E_0)}\nonumber\\
&&\qquad\qquad\qquad\qquad\quad
\times\rho_2^{-x_\phi-(L/2\pi)(E_\alpha-E_0)}\nonumber\\
&&\qquad\qquad\qquad\qquad\quad
\times\exp[i(L/2\pi)P_\alpha(\theta_2-\theta_1)]\; .
\label{e8}
\end{eqnarray}

When $\rho_2\gg\rho_1$ the leading contribution is given by the lowest excited
state $\vert\phi\rangle$ in the spectrum of the primary field $\phi$. According to
finite-size scaling theory,\cite{barber83} the matrix element scales as
\begin{eqnarray}
\langle0\vert\phi\vert\phi\rangle&=&f_\phi(\Delta_\chi,L)\nonumber\\
&=&L^{-x_\phi}f_\phi(L^{y_\chi}\Delta_\chi,1)\nonumber\\
&=&f_\phi[(2\pi)^{y_\chi}\alpha,1]L^{-x_\phi}\nonumber\\
&=&g(\alpha)L^{-x_\phi}\; ,
\label{e9}
\end{eqnarray}
whereas the mass gap is related to the correlation length universal scaling
function\cite{barber83,privman84}
\begin{eqnarray}
E_\phi-E_0=\xi_\phi^{-1}(\Delta_\chi,L)&=&[L\xi_\phi((2\pi)^{y_\chi}
\alpha,1)]^{-1}\nonumber\\
&=&(2\pi/L)x_\phi(\alpha)\; ,
\label{e10}
\end{eqnarray}
where we have introduced the perturbation-dependent quantity
\begin{equation}
x_\phi(\alpha)=[2\pi\xi_\phi((2\pi)^{y_\chi}\alpha,1)]^{-1}\; ,
\label{e11}
\end{equation}
which reduces to the scaling dimension of the primary field of the unperturbed
system when $\alpha=0$.\cite{cardy84} For a scalar field, the momentum $P_\phi$ of
the lowest excited state vanishes and the leading contribution to the two-point
function on the plane reads
\begin{eqnarray}
\langle\phi(\rho_1,\theta_1)\phi(\rho_2,\theta_2)\rangle\simeq
g^2(\alpha)\rho_1^{-x_\phi+x_\phi(\alpha)}&&
\rho_2^{-x_\phi-x_\phi(\alpha)}\nonumber\\
&&(\rho_2\gg\rho_1)\; ,
\label{e12}
\end{eqnarray}
so that $x_\phi(\alpha)$, given in Eq.~(\ref{e11}), may be interpreted as the
scaling dimension of $\phi$ near the defect.

In order to test these results, let us consider a perturbed Gaussian model at its
critical point, with a temperaturelike radial perturbation
\begin{equation}
\Delta_{\varphi^2}=\alpha\rho^{-2}\; ,
\label{e13}
\end{equation}
which is marginal since $y_{\varphi^2}=2$ in this case.\cite{toulouse75}
On the cylinder the free-energy functional is
\begin{equation}
F[\varphi]={\textstyle 1\over2}\int\!\! du\,
dv\,[(\nabla\varphi)^2+\alpha(2\pi/L)^2\varphi^2]\; ,
\label{e14}
\end{equation}
and the two-point correlation function
$\langle\varphi(0,0)\varphi(u,v)\rangle=g(u,v)$ satisfies the differential
equation\cite{bray77}
\begin{equation}
[-\nabla^2+\alpha(2\pi/L)^2]\, g(u,v)=\delta(u)\, \delta(v)\; ,
\label{e15}
\end{equation}
which is easily solved through a Fourier transformation leading to
\end{multicols}
\widetext
\noindent\rule{20.5pc}{.1mm}\rule{.1mm}{2mm}\hfill
\begin{equation}
g(u,v)=(1/4\pi)\sqrt{\alpha}\exp[-\sqrt{\alpha}(2\pi/L)u]+{1\over2\pi}
\sum_{n=1}^\infty1/(n^2+\alpha)^{1/2}\exp[-(n^2+\alpha)^{1/2}(2\pi/L)u]
\cos[n(2\pi/L)v]\; ,
\label{e16}
\end{equation}
so that $\alpha$ must be positive, a result which is related to the fact that
the low-temperature phase of the Gaussian model is ill defined.\cite{toulouse75}

On the plane the free-energy functional reads
\begin{equation}
F[\varphi]={\textstyle 1\over2}\int\!\!\rho\, d\rho\,
d\theta\,[(\nabla\varphi)^2+\alpha\rho^{-2}\varphi^2]\; ,
\label{e17}
\end{equation}
and the two-point correlation function satifies the equation
\begin{equation}
(-\nabla_1^2+\alpha\rho_1^{-2})\, g(\rho_2,\rho_1;\theta_2-\theta_1)=1/\rho_1\,
\delta(\rho_2-\rho_1)\,\delta(\theta_2-\theta_1)\; .
\label{e18}
\end{equation}
The solution is obtained as a linear combination of solutions of the homogeneous
equation and, with $\rho_2>\rho_1$, reads
\begin{equation}
g(\rho_2,\rho_1;\theta_2-\theta_1)=(1/4\pi)\sqrt{\alpha}
\rho_1^{\sqrt{\alpha}}\rho_2^{-\sqrt{\alpha}}+{1\over2\pi}
\sum_{n=1}^\infty1/(n^2+\alpha)^{1/2}\rho_1^{(n^2+\alpha)^{1/2}}
\rho_2^{-(n^2+\alpha)^{1/2}}\cos[n(\theta_2-\theta_1)]\; .
\label{e19}
\end{equation}
\hfill\rule[-2mm]{.1mm}{2mm}\rule{20.5pc}{.1mm}
\begin{multicols}{2} 
\narrowtext
A comparison with Eq.~(\ref{e12}) indicates that the defect exponent is
$x_\varphi(\alpha)=\sqrt{\alpha}$ since $x_\varphi=x_\varphi(0)=0$ for the
Gaussian model.\cite{toulouse75} This value is consistent with the expression of
the first gap in Eq.~(\ref{e16}). Furthermore, the two-point correlation functions
on the cylinder and on the plane, like the differential equations, transform into
one another under the change of variables given in Eq.~(\ref{e2}) as a consequence
of the conformal transformation given in Eq.~(\ref{e6}) with~$x_\varphi=0$.

For the Ising model, the thermal and magnetic defect exponents may be deduced
from the known scaling functions for the gaps of the corresponding primary
fields:\cite{hamer81,burkhardt87}
\begin{mathletters}
\label{e20}
\begin{eqnarray}
x_t(\alpha)&=&1+{\textstyle (c\alpha)^2\over2\pi^2}+O((c\alpha)^4)\; ,\\
x_h(\alpha)&=&{\textstyle
1\over8}+{\textstyle c\alpha\over4\pi}+{\textstyle\ln2\over4\pi^2}\,(c\alpha)^2
+O((c\alpha)^3)\; , 
\end{eqnarray}
\end{mathletters}
where $c$ is a lattice-dependent constant taking the value $8\pi$
(Ref.~\onlinecite{burkhardt87}) on the square lattice with isotropic couplings
for a radial perturbation $K(\rho)-K_c=-\alpha\rho^{-1}$.

More details and extensions to other geometries are planned to be published
elsewhere. Recently, a paper by R.Z.Bariev and I.Peschel was published [J.
Phys. A {\bf24}, L87 (1991)], in which the same conformal transformation is
used to study a radial marginal perturbation in the Ising model, either in the
bulk or at a surface. A phenomenological treatment of the critical behavior
near extended defects has also been given by Bariev (Zh. Eksp. Theor. Fiz.
{\bf94}, 374 (1988) [Sov. Phys. JETP {\bf67}, 2170 (1988)]).
\vfill\newpage
\acknowledgements The author is indebted to Ferenc Igl\'oi, Malte Henkel and
Bertrand Berche for useful discussions about these or related topics. The
Laboratoire de Physique du Solide is Unit\'e de Recherche Associ\'ee au CNRS
No. 155.

\end{multicols}

\end{document}